\documentclass[a4paper,11pt]{article}
\usepackage{paperstyle}

\usepackage{amsmath,amssymb,graphicx}
\usepackage{soul}
\usepackage[latin1]{inputenc}
\usepackage{dsfont}

\usepackage{subfigure}

\usepackage{array}
\usepackage{mathtools}

\usepackage[section, below, above]{placeins}

\usepackage[thinlines]{easytable}

\usepackage{color}

\newcommand{\be}{\begin{eqnarray}}
\newcommand{\en}{\end{eqnarray}}
\newcommand{\nn}{\nonumber\\}

\newcommand{\mpl}{m_{\rm{pl}}}
\newcommand{\As}{A_{\rm{s}}}

\usepackage{afterpage}

\begin{document}

\begin{titlepage}

\vspace*{-15mm}
\vspace*{0.7cm}

\begin{center}

{\Large {\bf Impact of other scalar fields on oscillons after hilltop inflation}}\\[8mm]

Stefan Antusch$^{\star\dagger}$\footnote{Email: \texttt{stefan.antusch@unibas.ch}},  
Stefano Orani$^{\star}$ \footnote{Email: \texttt{stefano.orani@unibas.ch}}

\end{center}

\vspace*{0.20cm}

\centerline{$^{\star}$ \it
Department of Physics, University of Basel,}
\centerline{\it
Klingelbergstr.\ 82, CH-4056 Basel, Switzerland}

\vspace*{0.4cm}

\centerline{$^{\dagger}$ \it
Max-Planck-Institut f\"ur Physik (Werner-Heisenberg-Institut),}
\centerline{\it
F\"ohringer Ring 6, D-80805 M\"unchen, Germany}

\vspace*{1.2cm}

\begin{abstract}
\noindent 
Oscillons are spatially localized and relatively stable field fluctuations which can form after inflation under suitable conditions. In order to reheat the universe, the fields which dominate the energy density after inflation have to couple to other degrees of freedom and finally produce the matter particles present in the universe today. In this study, we use lattice simulations in $2+1$ dimensions to investigate how such couplings can affect the formation and stability of oscillons.  We focus on models of hilltop inflation, where we have recently shown that hill crossing oscillons generically form, and consider the coupling to an additional scalar field which, depending on the value of the coupling parameter, can get resonantly enhanced from the inhomogeneous inflaton field. We find that three cases are realized: without a parametric resonance, the additional scalar field has no effects on the oscillons. For a fast and strong parametric resonance of the other scalar field, oscillons are strongly suppressed. For a delayed parametric resonance, on the other hand, the oscillons get imprinted on the other scalar field and their stability is even enhanced compared to the single-field oscillons.    

\end{abstract}
\end{titlepage}


\section{Introduction}

Inflation provides an attractive framework for explaining the initial conditions of hot big bang cosmology. During inflation, the universe undergoes a phase of accelerated expansion, driven by a dominating vacuum energy component. Its simplest realization consists of a single scalar field, the inflaton field, which is slowly rolling down a mild potential slope. The vacuum energy is then provided by the potential energy of the inflaton field, which, after inflation, is converted into radiation during the phase of reheating. 

Reheating is typically subdivided into two stages: an initial non-perturbative stage called preheating, and a final stage of perturbative particle decays and thermalization. Once thermal equilibrium is reached, the epoch of conventional hot big bang cosmology starts. Understanding the details of reheating is necessary in order to connect the inflationary dynamics to the subsequent evolution of the universe and finally to the underlying particle physics theory. Furthermore, to derive precise predictions of the primordial spectrum on CMB scales we need to know the exact expansion history of the universe during this phase.

Observations from the Planck satellite \cite{Ade:2015lrj,Ade:2015xua} constrain the possible shapes of the inflaton potential. Among the viable potentials \cite{Martin:2013nzq}, those with a plateau, such as in hilltop models \cite{Linde:1981mu,Izawa:1996dv,Senoguz:2004ky,Boubekeur:2005zm,Antusch:2013eca,Antusch:2014qqa}, are interesting candidates. 
In this class of models, inflation happens while the inflaton $\phi$ rolls away from the top of a hill and towards a minimum of its potential. Hilltop inflation of this type can be realized with the following inflaton potentials:
\begin{align}
 V(\phi) \, \simeq \, V_0 \left( 1 - \frac{\phi^p}{v^p} \right)^2 \,,
\label{eq:vhi}
\end{align}
where $p\geq2$ is an integer. With $p\geq6$, the model is compatible with the most recent $1\sigma$ Planck bounds.  Potentials of the form \eqref{eq:vhi} can be found in realistic particle physics models, where they are associated with, e.g.,   phase transitions breaking a GUT symmetry \cite{Linde:1981mu} or a flavour symmetry \cite{Antusch:2008gw}. 
The initial conditions for hilltop inflation models can be obtained with a preinflation mechanism, by which the position of the inflaton at the top of the hill is dynamically generated. This motivates the introduction of a second scalar field $\chi$, which forces the inflaton to be at the maximum of its potential (see, for example, \cite{Senoguz:2004ky,Izawa:1997df,Antusch:2014qqa}). 

The introduction of other fields besides the inflaton is of course also necessary in order to reheat the universe, since, as already mentioned above, the inflaton has to transfer its energy to other degrees of freedom in order to finally produce the matter particles of the present universe.  An interesting possibility in this context is to indentify the scalar field $\chi$ introduced as a preinflaton in \cite{Antusch:2014qqa} with a right-handed sneutrino. Via the decays of the sneutrinos into Higgs(inos) and (s)leptons, the universe can reheat efficiently and at the same time the matter-antimatter asymmetry may be generated via non-thermal leptogenesis.    

Preheating after hilltop inflation is challenging due to the non-linear nature of the field dynamics: as the inflaton rolls down the potential and oscillates around the minimum, its fluctuations get enhanced by ``tachonic preheating'' as well as ``tachyonic oscillations'' \cite{Desroche:2005yt,Brax:2010ai,Antusch:2015nla} and quickly become non-linear. As has recently been shown in \cite{inprep}, the highly inhomogeneous inflaton field can furthermore excite other scalar fields, such as the $\chi$ field, via parametric resonance. Whether this resonant enhancement of the $\chi$ fluctuations occurs depends on the value of the coupling between $\phi$ and $\chi$.

A particularly interesting phenomenon which can arise during preheating when the potential is shallower than quadratic away from the minimum \cite{Amin:2011hj}, is the formation of localized oscillating configurations of the inflaton field $\phi$, called oscillons \cite{Copeland:1995fq,Amin:2011hj}.  After hilltop inflation with potential of the form of eq.~\eqref{eq:vhi}, it has recently been shown that ``hill crossing'' oscillons generically form, with the field initially oscillating between the two minima of the potential \cite{Antusch:2015nla}. Eventually, the hill crossing stops (i.e.\ the bubbles of ``wrong vacuum'' finally collapse and do not reform) and the oscillons fluctuate around the ``true'' minimum towards which the inflaton was initially rolling.

Oscillons are of course not unique to hilltop models: other inflationary models where oscillons have been shown to form include hybrid inflation \cite{Copeland:2002ku} and supersymmetric extensions of it \cite{McDonald:2001iv,Broadhead:2005hn}. In addition to inflation, oscillons can also arise in high-energy physics, such as Abelian Higgs models \cite{Gleiser:2007te,Achilleos:2013zpa} and the Standard Model of particle physics \cite{Farhi:2005rz,Graham:2006vy}. They naturally arise in field theories as a remnant of collapse and collisions of false vacuum bubbles \cite{Copeland:1995fq,Bond:2015zfa}. Their role during preheating has been studied using lattice simulations in a variety of inflationary models and they are known to have a suprisingly long lifetime \cite{Graham:2006xs,Gleiser:2009ys,Amin:2010jq,Gleiser:2014ipa}, although quantum corrections can accelerate their decay \cite{Saffin:2014yka}. Possible observational consequences of oscillons range from the production of the primordial baryon asymmetry \cite{Lozanov:2014zfa} to the generation of gravitational waves during their formation and decay \cite{Zhou:2013tsa}. Furthermore, they can delay thermalization \cite{Gleiser:2006te} thus having an effect on the primordial curvature perturbation. The impact of couplings to other scalar fields on oscillons after hilltop inflation has been investigated in \cite{Gleiser:2014ipa} for a potential similar to eq.~\eqref{eq:vhi}, however no effects have been observed.

In this paper we revisit the question whether couplings between the inflaton and other fields can affect the formation and stability of oscillons. We consider hilltop inflation with a potential as in eq.~\eqref{eq:vhi} with $p=6$ and focus on the coupling to an additional scalar field $\chi$ which, as discussed in \cite{inprep}, can get resonantly enhanced from the inhomogeneous inflation field. To this end, we use lattice simulations in $2+1$ dimensions and show that, depending on the strength of the coupling between $\phi$ and $\chi$, and therefore on the occurrence and timing of the parametric resonance of $\chi$, the formation and stability of the oscillons can be either enhanced, suppressed or left unaffected.  We expect the resonant amplification of $\chi$ caused by the inhomogeneous inflaton to also have a strong impact on oscillons in $3+1$ dimensions.

The paper is structured as follows: in section~\ref{sec:hilltopModel}, we introduce the hilltop inflation model under consideration and summarize its predictions for the inflationary observables. A short overview of the dynamics of preheating after hilltop inflation is given in section~\ref{sec:preh}. In section~\ref{sec:osc}, we introduce the method used for the lattice simulations in $2+1$ dimensions and present our results. Finally, in section~\ref{sec:conc}, we conclude with a summary and discussion of the results.

\section{Hilltop inflation}
\label{sec:hilltopModel}

The potential $V(\phi,\chi)$ we consider for our analysis contains the hilltop inflation potential of eq.~\eqref{eq:vhi} with $p=6$ plus an interaction term between the inflaton $\phi$ and the additional scalar field $\chi$:
\begin{align}
 V(\phi,\chi) \, = \, V_0 \left( 1 - \frac{\phi^6}{v^6} \right)^2 + \frac{\lambda^2}{2}\phi^2\chi^2\left(\phi^2+\chi^2\right)\,,
\label{eq:potential}
\end{align}
where $v\propto\mpl$, $V_0\propto\mpl^4$  and $\lambda\propto\mpl^{-1}$.

\vspace{1mm}
\subsubsection*{Observable inflation: initial conditions and predictions}

For the hilltop inflation model of eq.~(\ref{eq:potential}), the initial conditions for observable inflation are generated dynamically via a preinflation mechanism, as shown in \cite{Antusch:2014qqa}. Starting with large field values of $\phi$ and $\chi$, a mass for $\phi$ is generated from the interaction term, which efficiently drives $\phi$ towards $0$. Subsequently, $\chi$ is slowly rolling towards $0$ during a phase of preinflation. With both $\phi$ and $\chi$ close to $0$, the induced mass term for $\phi$ becomes so small that the fields are in a diffusion region where quantum fluctuations dominate over classical rolling. When the fields eventually leave the diffusion region and $\phi$ starts rolling down the top of the hill towards one of the minima of the potential at $\phi=\pm v$ (while $\chi \approx 0$), this provides suitable initial conditions for the observable phase of inflation to start. As discussed in \cite{Antusch:2014qqa,inprep}, the value of $\chi$ at this stage is typically very small and we set $\chi=0$ at the start of observable inflation in what follows. Furthermore, without loss of generality, we choose to roll along $\phi>0$.

Under these conditions, the potential during observable inflation simplifies to

\begin{align}
 V(\phi) \, \simeq \, V_0 \left( 1 - \frac{\phi^6}{v^6} \right)^2 \,.
\label{eq:vinf}
\end{align}
In order to solve the flatness and horizon problems, this phase of inflation lasts more than $N_*\equiv\ln(a_e/a_*)\sim 60$ $e$-folds, where $a$ is the scale factor, the subscripts $e$ and $*$ denote, respectively, the end of inflation and the time when perturbations on the scales of the CMB exit the horizon. Inflation ends when $\phi$ exits the slow-roll regime and accelerates towards the minimum. 

The scalar spectral index $n_{\rm s}$ and the tensor to scalar ratio $r$ of the CMB perturbations are given in terms of the slow-roll parameters, $\epsilon_\phi\equiv1/2\mpl^2\left(\partial V/\partial\phi\right)^2/V^2$ and $\eta_\phi\equiv\mpl^2\left(\partial^2V/\partial\phi^2\right)/V$, evaluated at $N_*$:
\be
n_{\rm s}\, &=& \, 1-6\epsilon_\phi(\phi_*)+2\eta_\phi(\phi_*) \simeq 1- \frac{10}{5+4N_*} \simeq 0.96\,,\nn
r\, &=& \, 16\epsilon_\phi(\phi_*) \simeq 4\times10^{-6}\left(\frac{v}{\mpl}\right)^3\,,
\label{eq:nsr}
\en
where $\phi_*/\mpl\simeq 0.1 (v/\mpl)^{3/2}$.
The predictions for $v<\mpl$ are compatible with the most recent Planck bounds $n_{\rm s}=0.968\pm0.006$ at $68\%$ CL and $r<0.09$ at $95\%$ CL \cite{Ade:2015lrj,Ade:2015xua}. Furthermore, the amplitude of CMB scalar fluctuations, $A_{\rm s}\simeq2.2\times10^{-9}$, leads us to fix the value of $V_0$ to 
\be
V_0 \, = \, 24\pi^2\varepsilon_\phi({\phi_*})\As\,m_{\rm Pl}^4 \, \simeq \, 10^{-13}\,v^3\,m_{\rm{Pl}}\,.
\label{eq:V0}
\en

\vspace{1mm}
\subsubsection*{Derivation of the potential from supersymmetry}

The potential of eq.~\eqref{eq:potential} can be derived from supersymmetry. Its form follows from the superpotential
\be
W\,=\,\sqrt{V_0}\hat{S}\left(\frac{8\hat{\Phi}^6}{v^6} - 1  \right) + \lambda\hat{\Phi}^2\hat{X}^2\,,
\label{eq:sp}
\en
where $\hat{\Phi}$ and $\hat{X}$ are, respectively, the chiral superfields containing $\phi = \sqrt{2} \mbox{Re[$\Phi$]}$ and $\chi = \sqrt{2} \mbox{Re[$X$]}$, the real scalar fields which appear in eq.~\eqref{eq:potential}. As discussed in \cite{inprep}, we set the imaginary components of the fields to zero. The imaginary component of the inflaton can affect the dynamics, leading to different inflationary dynamics and prediction for the primordial spectrum \cite{Nolde:2013bha}. However, for large part of parameter space, the inflationary dynamics reduce to the single-field limit. The imaginary part of $X$ has the same equation of motion as the real part $\chi$, leading us, for simplicity, to consider only $\chi$. In addition, we assume that the scalar field $S$ contained in $\hat{S}$ is initially confined at $S=0$ due to a K\"ahler induced super-Hubble mass during inflation \cite{Antusch:2008pn}. As we have shown in \cite{inprep}, even if $S=0$ initially, its fluctuations are efficiently enhanced during preheating, such that $S$ essentially evolves as $\phi$, without significantly affecting the behaviour of $\chi$. Based on this, in what follows we do not include $S$. Moreover, we assume negligible K\"ahler corrections to the masses of $\phi$ and $\chi$. This can either be by accident or due to symmetry (e.g.\ a Heisenberg symmetry \cite{Antusch:2013eca}). Therefore, for the simulations in this paper, we will restrict ourselves to the potential of eq.~\eqref{eq:potential}.
 
The advantage of the suspersymmetric formulation of the model is that the form of the superpotential~\eqref{eq:sp} can be justified by imposing a $U(1)_{\rm R}$ and a $\mathbb{Z}_p$ symmetry. Then, $\hat{S}$ has two units of $U(1)_{\rm R}$ charge and is a $\mathbb{Z}_p$-singlet. $\hat{\Phi}$ is a $U(1)_{\rm R}$-singlet and has one unit of $\mathbb{Z}_p$ charge. Finally, the form of the interaction term is fixed by giving $\hat{X}$ one unit of $U(1)_{\rm R}$ charge and two units of $\mathbb{Z}_p$ charge.

\section{Preheating dynamics after Hilltop inflation: overview}
\label{sec:preh}

In this section we give a brief overview of preheating after hilltop inflation models of the type of eq.~\eqref{eq:potential}. The dynamics outlined here are discussed in greater details in \cite{Antusch:2015nla} and \cite{inprep}. It is convenient for what follows to consider the Fourier decomposition of a field $f$ in ${\rm D}$ spatial dimensions: $f\left(t,\vec{x}\right)=\int\frac{d^{\rm D}k}{(2\pi)^{\rm D}}f_k(t)e^{-i\vec{k}\cdot\vec{x}}$, where $k\equiv|\vec{k}|$. We will first describe the dynamics of preheating after inflation for the single-field scenario and then discuss how they are affected when $\lambda \not=  0$.

As the inflaton leaves the slow-roll regime and accelerates towards $\phi=v$,  {\it tachyonic preheating} very efficiently amplifies modes $k$ for which $k^2/a^2 + \partial^2V/\partial\phi^2 < 0$. The efficiency of this amplification grows as $v$ decreases and for $v\lesssim10^{-5}\mpl$, the system becomes non-linear already during this phase.

\begin{figure}[tbp]
  \centering$
\begin{array}{cc}
\includegraphics[width=0.48\textwidth]{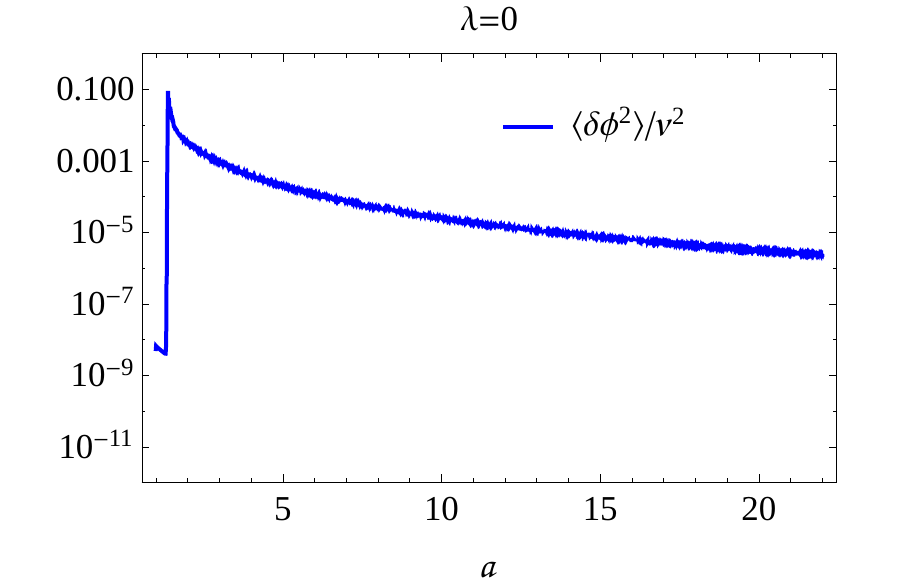}
\includegraphics[width=0.48\textwidth]{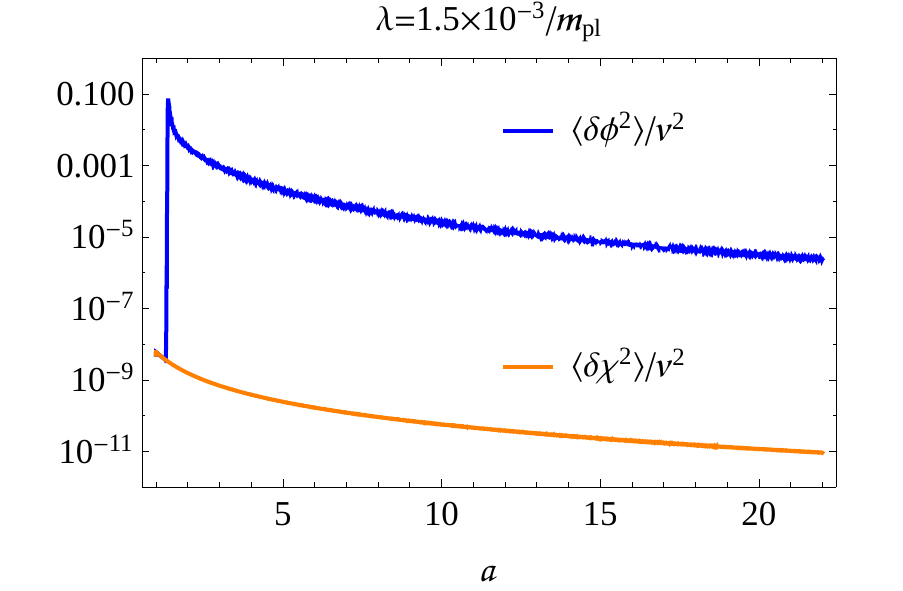}\\
\includegraphics[width=0.48\textwidth]{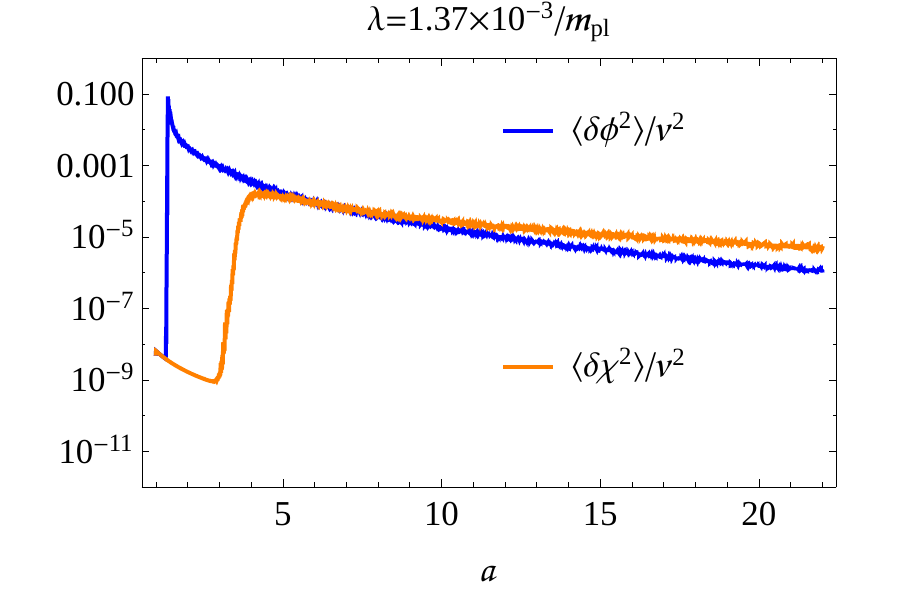}
\includegraphics[width=0.48\textwidth]{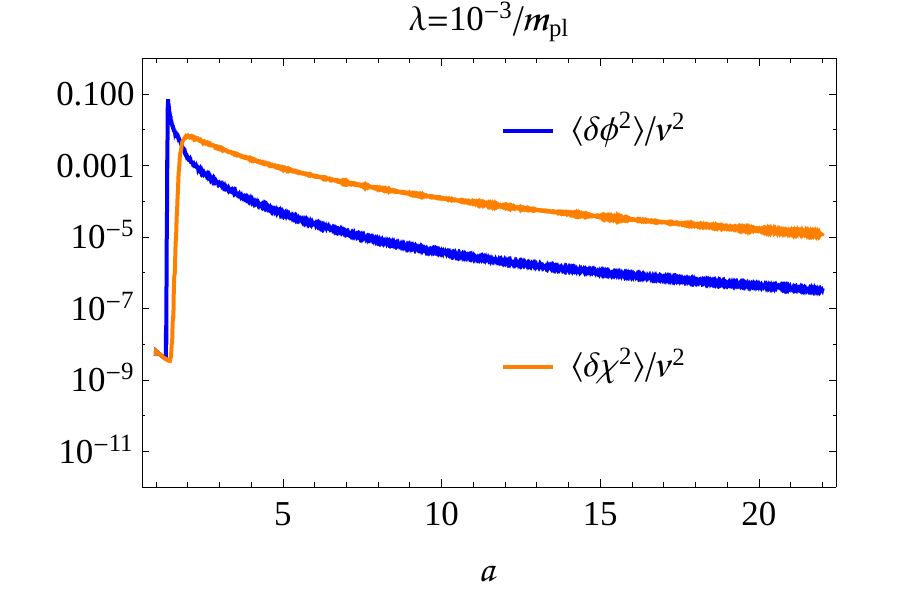}
\end{array}$
 \caption{Variances $\langle\delta\phi^2\rangle/v^2$ and $\langle\delta\chi^2\rangle/v^2$ for $v=10^{-2}\mpl$ and $\lambda=0$ (top left), $\lambda=1.5\times10^{-3}/\mpl$ (top right), $\lambda=1.37\times10^{-3}/\mpl$ (bottom left) and $\lambda=10^{-3}/\mpl$ (bottom right) from simulations in $2+1$ dimensions with $1024^2$ points. The behaviour of $\langle\delta\phi^2\rangle$ is qualitatively similar in the four plots: the $\phi$ fluctuations grow rapidly when $\phi$ oscillates around $v$ until its homogeneous mode decays because of non-linear interactions at $a\sim1.5$. On the other hand, $\chi$'s behaviour depends on the value of $\lambda$. For $\lambda=1.5\times10^{-3}/\mpl$ the amplitude of its perturbations redshifts from the initial vacuum fluctuations.
For $\lambda=1.37\times10^{-3}/\mpl$ and $\lambda=10^{-3}/\mpl$, the amplitude of $\chi$'s fluctuations eventually grows due to a parametric resonance with the inhomogeneous inflaton field \cite{inprep}. The timing and efficiency of the resonance depends on the strength of the coupling: for $\lambda=1.37\times10^{-3}/\mpl$, $\langle\delta\chi^2\rangle\sim\langle\delta\phi^2\rangle$ at $a\sim4$, whereas for $\lambda=10^{-3}/\mpl$, $\chi$'s variance grows earlier, exceeding $\phi$'s at $a\sim2$ and settling at $\langle\delta\chi^2\rangle\sim10\langle\delta\phi^2\rangle$.}
  \label{fig:vars}
\end{figure}

For $v\gtrsim10^{-5}\mpl$, this phase is followed by a period of {\it tachyonic oscillations} of $\phi$ around $\phi=v$, when the homogeneous $\phi$ periodically enters the tachyonic region where $\partial^2V/\partial\phi^2 < 0$. Exponential growth and Hubble damping of the amplitude of the fluctuations lead to a net growth of the inflaton's modes peaked at $k_{\rm peak}\sim500\sqrt{V_0/3}/\mpl$.  For $v\gtrsim10^{-1}\mpl$, Hubble damping rapidly prevents the inflaton from entering the tachyonic region, stopping the growth of the field's fluctuations before they become non-linear. However, for $10^{-1}\gtrsim v/\mpl \gtrsim 10^{-5}$, the fluctuations become large enough so that $\langle\delta\phi^2\rangle\sim 0.1 v^2$, see fig.~\ref{fig:vars}. This growth leads to the formation of localized field fluctuations, i.e.\ oscillons, which initially oscillate between $-v\lesssim\phi\lesssim v$ and eventually settle at the initial minimum at $\phi=v$ \cite{Antusch:2015nla}. These oscillons are separated by a characteristic distance $d_{\rm osc} \simeq 2\pi/k_{\rm peak}$.

In \cite{inprep}, it has been shown that a non-zero coupling $\lambda$ between the inflaton $\phi$ and another scalar field $\chi$ can lead to a parametric resonance of $\chi$ after $\phi$ has become inhomogeneous and oscillons have formed. In particular, for $v=10^{-2}\mpl$ there is a resonance band at values of $\lambda$ such that the ratio of the mass of $\chi$ to the mass of $\phi$ at the global minimum, $m_\chi/m_\phi$, is around (and somewhat below) $0.5$. With
\be
m^2_{\phi} \, \equiv \, \left.\frac{\partial^2 V}{\partial \phi^2}\right|_{\rm{min}} \, = \,\frac{72 V_0}{v^2}\,, \quad 
m^2_{\chi} \, \equiv \, \left.\frac{\partial^2 V}{\partial \chi^2}\right|_{\rm{min}} \, = \,\lambda^2\,v^4\, \Rightarrow \quad
\frac{m_{\chi}}{m_{\phi}} \, = \, \frac{\lambda\,v^3}{\sqrt{72\,V_0}}\,,
\en
the resonance band has been found in \cite{inprep} to lie in the range $0.25 \lesssim  m_\chi/m_\phi \lesssim 0.5$, which corresponds to $0.7 \times 10^{-3}/m_\mathrm{pl} \lesssim  \lambda \lesssim 1.4 \times 10^{-3}/m_\mathrm{pl} $ 
This resonance can be very efficient, leading to $\langle\delta\chi^2\rangle\gtrsim\langle\delta\phi^2\rangle$ for certain $\lambda$s, see fig.~\ref{fig:vars}. When the parametric resonance occurs, the fluctuations of $\chi$ are peaked at scales close to $k_{\rm peak}$.

Motivated by these results, in the remainder of this paper we study the formation and evolution of the oscillons in the hilltop inflation model of eq.~\eqref{eq:potential},  choosing $v=10^{-2}\mpl$ as an example value, as in \cite{inprep}. We will compare the dynamics of the single-field case $\lambda=0$ to three different choices of $\lambda\neq0$ (see fig.~\ref{fig:vars}): a value that lies outside the resonance band, for which $\chi$'s fluctuations redshift from their initial vacuum amplitude; and two other values that lie inside the resonance band and which differ regarding the timing and the strength of the parametric resonance, which, nevertheless, occurs in both cases after the inflaton has become inhomogeneous. We will see in section \ref{sec:osc} that a coupling outside the resonance band has no significant effect on the evolution of oscillons. 
On the other hand, values of $\lambda$ inside the resonance band have a dramatic effect: depending on the time at which the fluctuations of $\chi$ grow, the oscillons can be either enhanced or suppressed.

\section{Evolution of oscillons}
\label{sec:osc}

In this section we present and analyse the results of lattice simulations of the hilltop inflation model of eq.~\eqref{eq:potential}. The aim is to understand the evolution of the oscillons that form at the end of inflation.
To this end, we numerically solve the non-linear equations of motion
\begin{align}
&\ddot{f}(t,\bar{x}) + 3H\dot{f}(t,\bar{x}) - \frac{1}{a^2}\bar{\nabla}^2f(t,\bar{x}) + \frac{\partial V}{\partial f} \, = \, 0\,, \label{eq:nleom1}\\ 
&H^2\, \equiv\, \frac{\langle\rho\rangle}{3\mpl^2}\,=\,\frac{1}{3\mpl^2}\left\langle V  +  \sum_f\left(\frac{1}{2}\dot{f}^2 + \frac{1}{2a^2} \left|\bar{\nabla}f\right|^2 \right)\right\rangle\,, 
\label{eq:nleom}
\end{align}
where $f$ represents the fields, $\phi$ and $\chi$, $\bar{\nabla}$ is the gradient with respect to the comoving coordinates $\bar{x}$ and $\langle .. \rangle$ denotes the average over space. We use a modified version of the program LATTICEEASY \cite{Felder:2000hq} to solve these equations on a discrete spacetime lattice.

 Since oscillons are known to remain stable over a surprisingly large number of oscillations, the lattice simulations have to run for long timescales. This leads us to focus on $2+1$ dimensions with $1024^2$ points. 

\subsection{Lattice initialization}
\label{sec:ini}

\begin{table}[tbp]
\centering
\begin{tabular}[c]{ | c | c | c |c | c | c | c |}
\hline
$v/\mpl$ & $\langle\phi\rangle_{\rm i}/v$ & \vphantom{$\frac{\dot{f}}{\dot{f}}$}$\langle\dot{\phi}\rangle_{\rm i}/v^2$ & $\langle\chi\rangle_{\rm i}/v$ & $\langle\dot{\chi}\rangle_{\rm i}/v^2$ & $H_{\rm i}/\mpl$ \\
\hline
\vphantom{$\frac{f}{f}$} $10^{-2}$ & $0.08$ & $2.49\times10^{-9}$ & 0 & 0 & $1.9\times10^{-10}$ \\ 
\hline \hline
D & $N$ & $k_{\rm uv}$ & $k_{\rm ir}$ & $\delta x$ & $L$ \\ 
\hline
 $2$ & $1024$ & $6.2\times10^4H_{\rm i}$ & $60.5H_{\rm i}$ & $10^{-4}/H_{\rm i}$ & $0.1/H_{\rm i}$ \\
\hline 
\end{tabular}
\caption{ Initial values of $\langle\phi\rangle$ and $\langle\dot{\phi}\rangle$ and other parameters of the 2D lattice simulations.}
\label{tab:ini}
\end{table}

We start the lattice simulations shortly after the end of inflation, when $|\eta_{\phi}|>1$ but $\epsilon_{\phi}<1$. At this time the system is still linear and the scalar fields' fluctuations are well described by the vacuum spectrum. Initial conditions and parameters are shown in table \ref{tab:ini}. In what follows, we restrict ourselves to $\chi=0$ at the end of inflation. However, $\chi$ can acquire a non-zero value at the end of inflation and its main effect on the preheating dynamics is to increase the efficiency of the parametric resonance (see \cite{inprep}).

The fields are initialized in Fourier space as random fields whose norms obey the Rayleigh distribution with variance given by the vacuum fluctuations:
\be
f_k &=& \frac{1}{a}\frac{|f_k|}{\sqrt{\alpha_{+}^2+\alpha_{-}^2}}\left(\alpha_{+}e^{i2\pi\theta_+ +i k t} + \alpha_{-}e^{i2\pi\theta_- -i k t}\right)\,,\nn
\dot{f}_k &=& \frac{i k}{a}\frac{|f_k|}{\sqrt{\alpha_{+}^2+\alpha_{-}^2}}\left(\alpha_{+}e^{i2\pi\theta_+ +i k t} - \alpha_{-}e^{i2\pi\theta_- -i k t}\right) -Hf_k\,,
\label{eq:fk}
\en
where $\alpha_+$, $\alpha_-$, $\theta_+$ and $\theta_-$ are random real numbers uniformly distributed between $0$ and $1$.

The fields \eqref{eq:fk} are then Fourier transformed to position space, where a discretized version of eqs.~\eqref{eq:nleom} is solved with lattice spacing $\delta x = 2\pi/k_{\rm uv}$, where $k_{\rm uv}$ is the ultraviolet cutoff. The infrared cutoff is given by $k_{\rm ir} = k_{\rm uv}/N$, where $N$ is the number of points per spatial dimension. The lattice length is $L=2\pi/k_{\rm ir}$.

In subsection \ref{sec:numre} we present the numerical results of lattice simulations in 2 spatial dimensions with parameters given in table \ref{tab:ini}. We study the evolution of the single-field case, corresponding to $\lambda=0$ in eq.~\eqref{eq:potential}, and of the two-field case with  $\lambda=1.5\times10^{-3}/\mpl$,  $\lambda=1.37\times10^{-3}/\mpl$ and $\lambda=10^{-3}/\mpl$.

\subsection{Numerical results}
\label{sec:numre}


\begin{figure}[tbp]
  \centering$
\begin{array}{cc}
\includegraphics[width=0.4\textwidth]{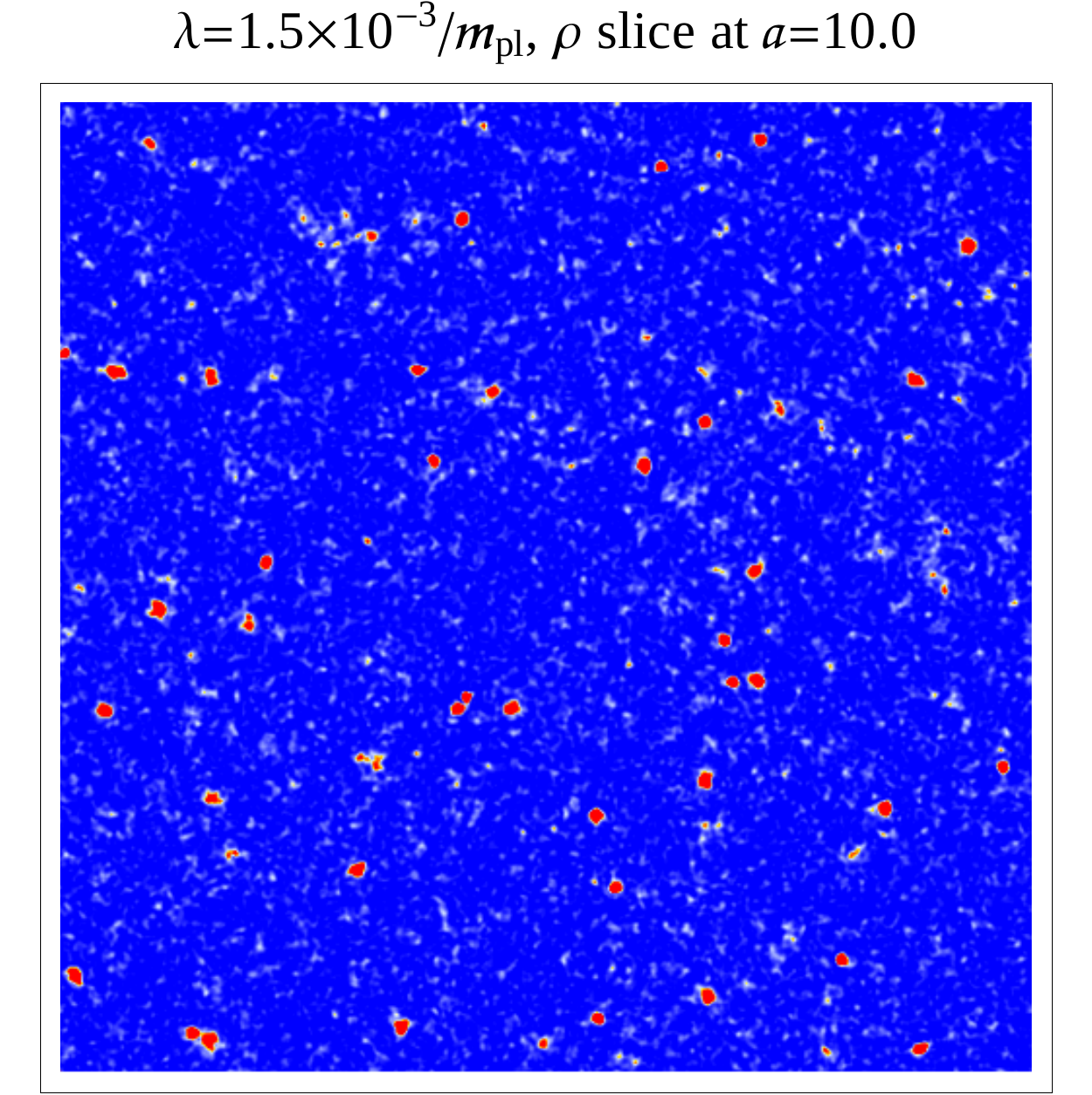}\;\;
\includegraphics[height=6.5cm]{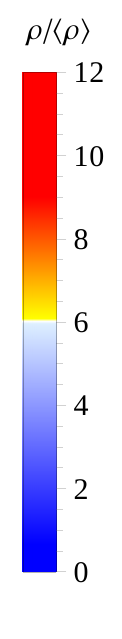}
\includegraphics[width=0.4\textwidth]{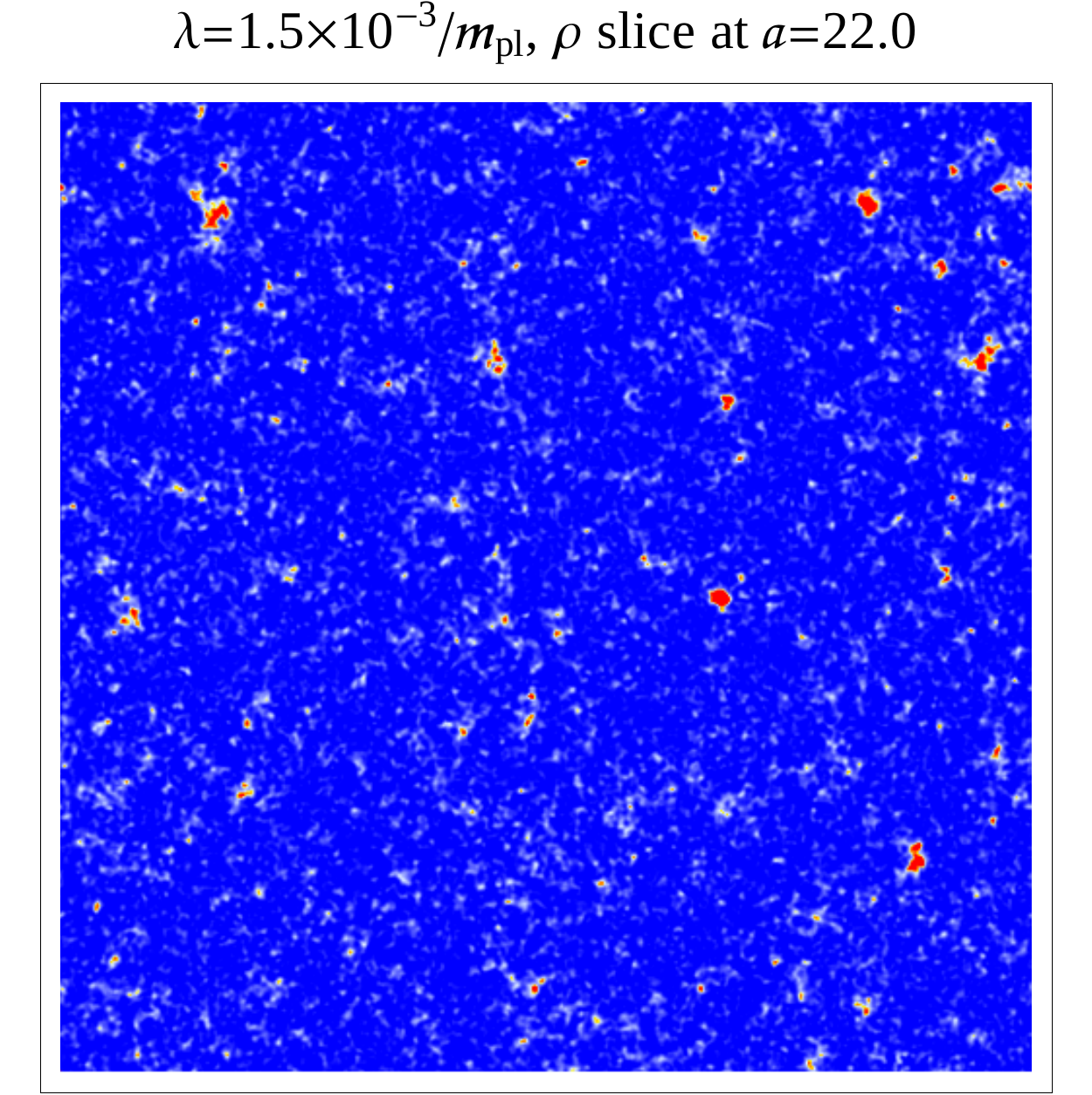}\\
\includegraphics[width=0.85\textwidth]{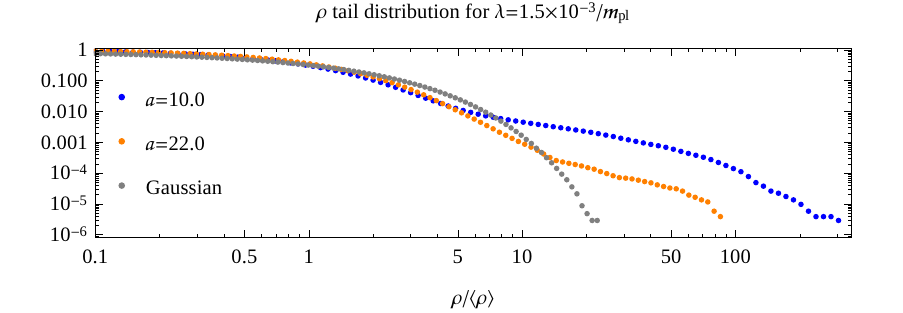}
\end{array}$
 \caption{{\it Above}: energy density slices at $a=10$ (left) and $a=22$ (right) for $v=10^{-2}\mpl$ and $\lambda=1.5\times10^{-3}/\mpl$, from a simulation with $1024^2$ points. Oscillons (localized red spots) are clearly visible at $a=10$ but most of them have decayed by $a=22$. The single-field case $\lambda=0$ is qualitatively the same. \newline
{\it Below}: tail distribution of the energy density $\rho$ as a function of $\rho/\langle \rho \rangle$ at $a=10$ (blue) and $a=22$ (orange) for $\lambda=1.5\times10^{-3}/\mpl$. In grey we show the tail distribution of $\rho_g = 36V_0 g_1^2/v^2 + \lambda^2 v^4g_2^2/2$ where $g_1$ and $g_2$ are two discrete Gaussian random fields defined on a lattice with $1024^2$ points, with  zero means and standard deviations chosen such that $\langle g_1^2\rangle/\langle g_2^2 \rangle \sim \langle\delta\phi^2\rangle/\langle\delta\chi^2\rangle\sim10^6$. The largest $\rho$ goes from $\rho>300\langle\rho\rangle$ at $a=10$ to $\rho<100\langle\rho\rangle$ at $a=22$. Furthermore, at $a=22$ the tail distribution is closer to what one would expect from a Gaussian random field. This confirms that most oscillons decay by the end of the lattice simulation. For the single-field case $\lambda=0$ we obtain qualitatively the same results.}
  \label{fig:ehist1}
\end{figure}

\begin{figure}[tbp]
  \centering$
\begin{array}{cc}
\includegraphics[width=0.4\textwidth]{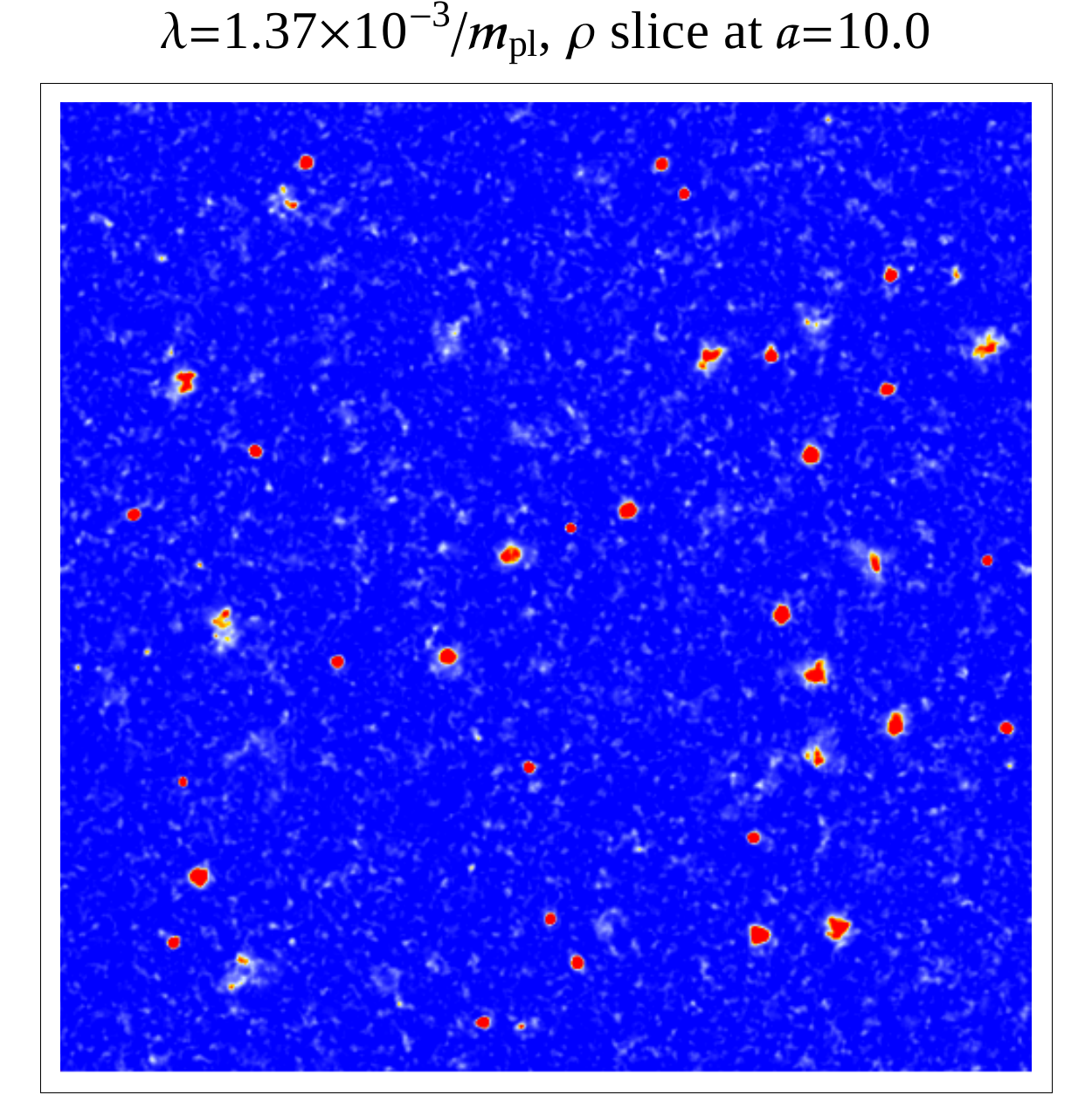}\;\;
\includegraphics[height=6.5cm]{graphics/rholegend}
\includegraphics[width=0.4\textwidth]{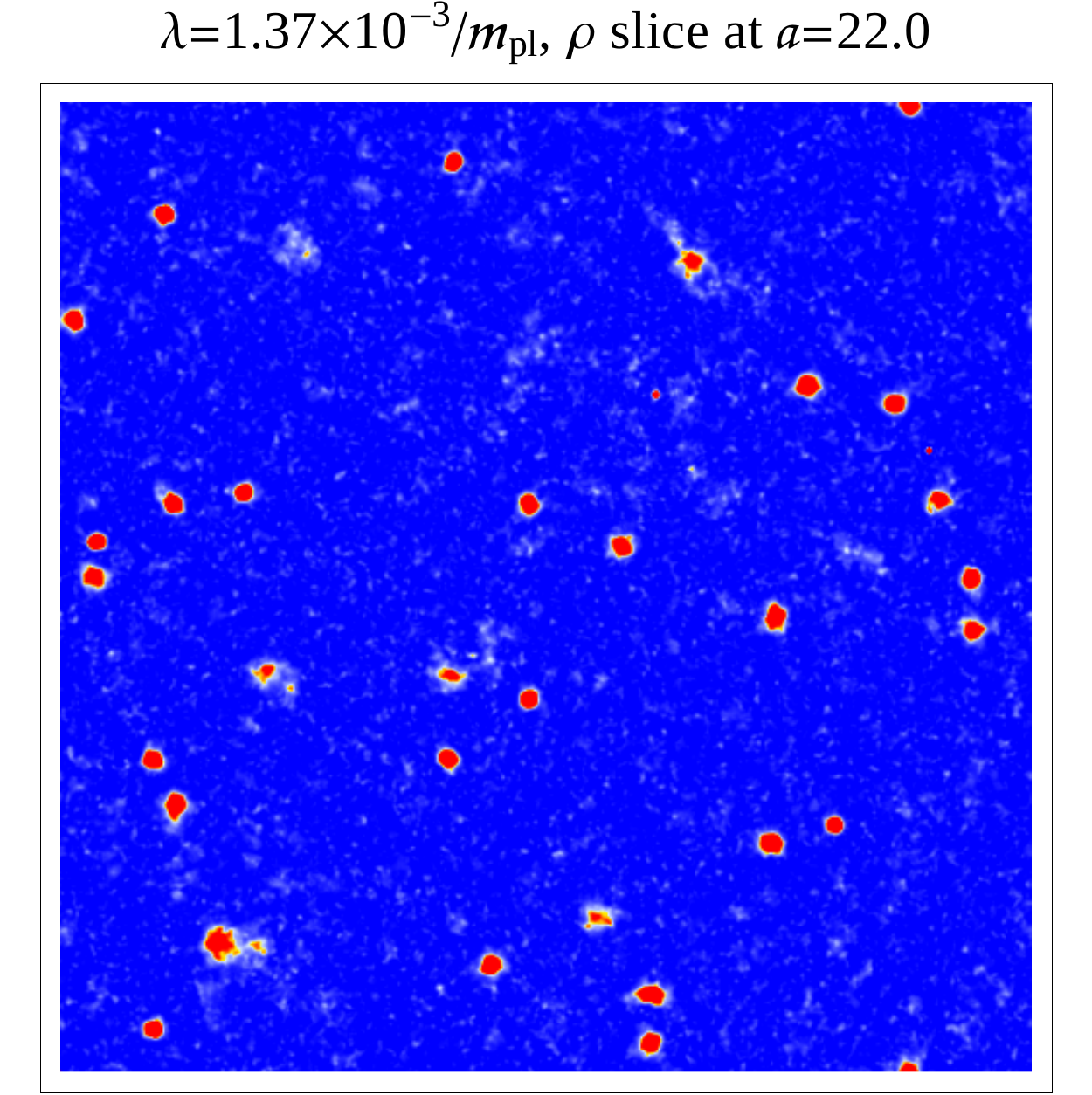}\\
\includegraphics[width=0.85\textwidth]{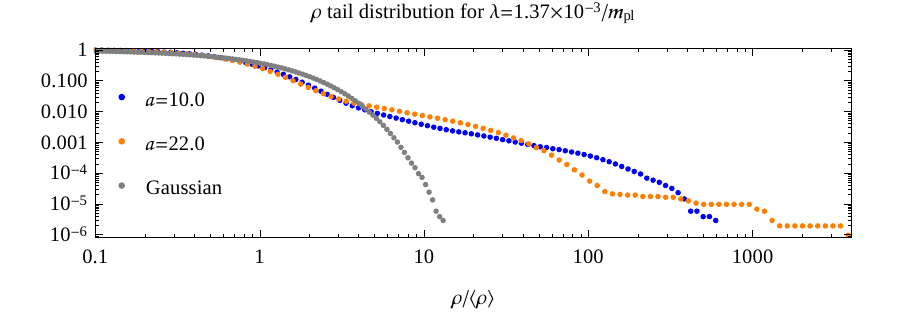}
\end{array}$
 \caption{{\it Above}: energy density slices at $a=10$ (left) and $a=22$ (right) for $v=10^{-2}\mpl$ and $\lambda=1.37\times10^{-3}/\mpl$, from a simulation with $1024^2$ points. Oscillons are clearly visible both at $a=10$ and $a=22$. \newline
{\it Below}: tail distribution of the energy density $\rho$ as a function of $\rho/\langle \rho \rangle$ at $a=10$ (blue) and $a=22$ (orange) for $\lambda=1.37\times10^{-3}/\mpl$. In grey we show the tail distribution of $\rho_g = 36V_0 g_1^2/v^2 + \lambda^2 v^4g_2^2/2$ where $g_1$ and $g_2$ are two discrete Gaussian random fields defined on a lattice with $1024^2$ points, with  zero means and standard deviations chosen such that $\langle g_1^2\rangle/\langle g_2^2 \rangle \sim \langle\delta\phi^2\rangle/\langle\delta\chi^2\rangle\sim1/2$. Both $a=10$ and $a=22$ tail distributions spread to $\mathcal{O}(1000)\langle\rho\rangle$, far from the Gaussian expectation, indicating that oscillons are enhanced compared to the non-resonant case $\chi$ in fig.~\ref{fig:ehist1}. }
  \label{fig:ehist2}
\end{figure}

\begin{figure}[tbp]
  \centering$
\begin{array}{cc}
\includegraphics[width=0.4\textwidth]{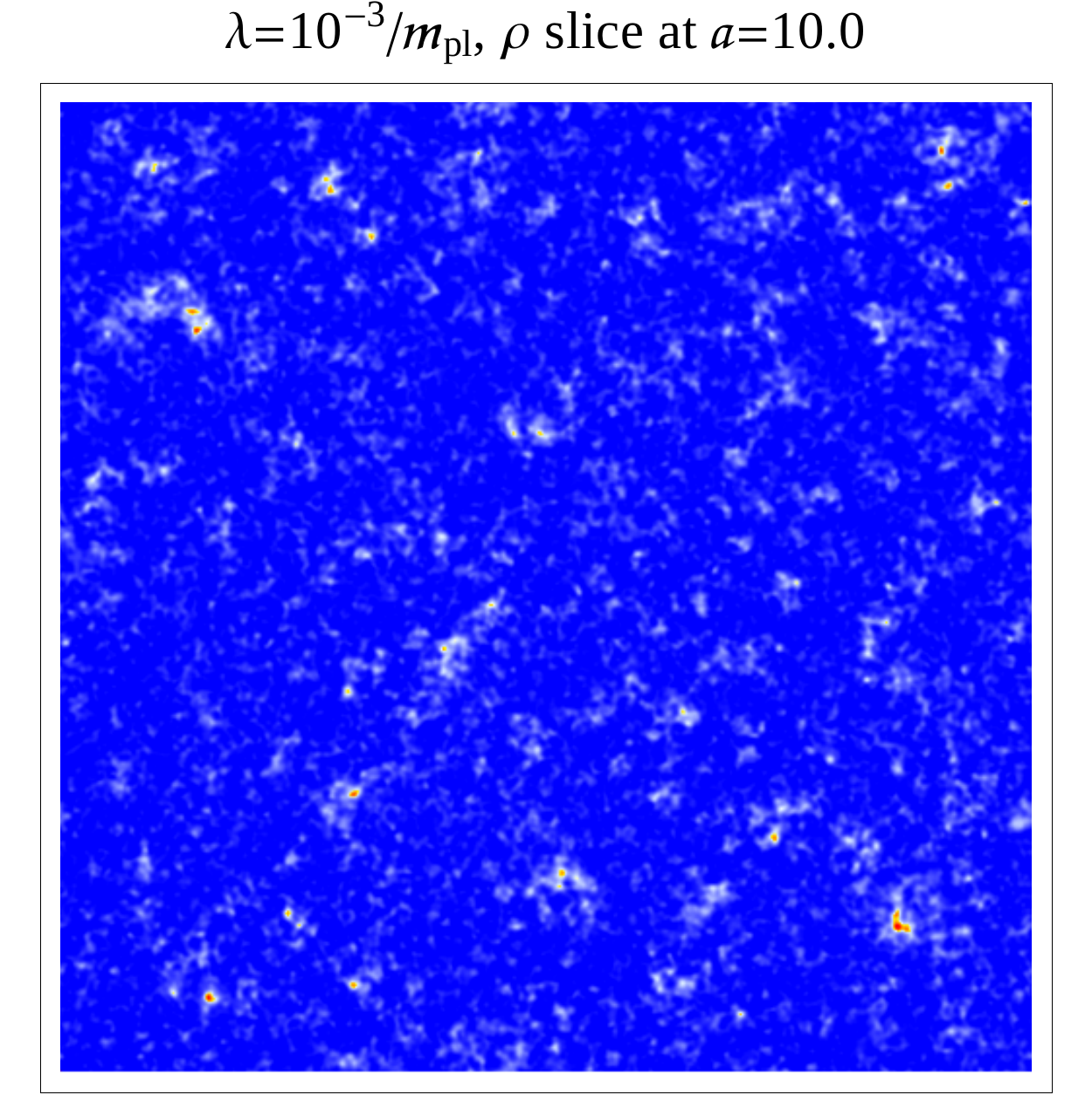}\;\;
\includegraphics[height=6.5cm]{graphics/rholegend}
\includegraphics[width=0.4\textwidth]{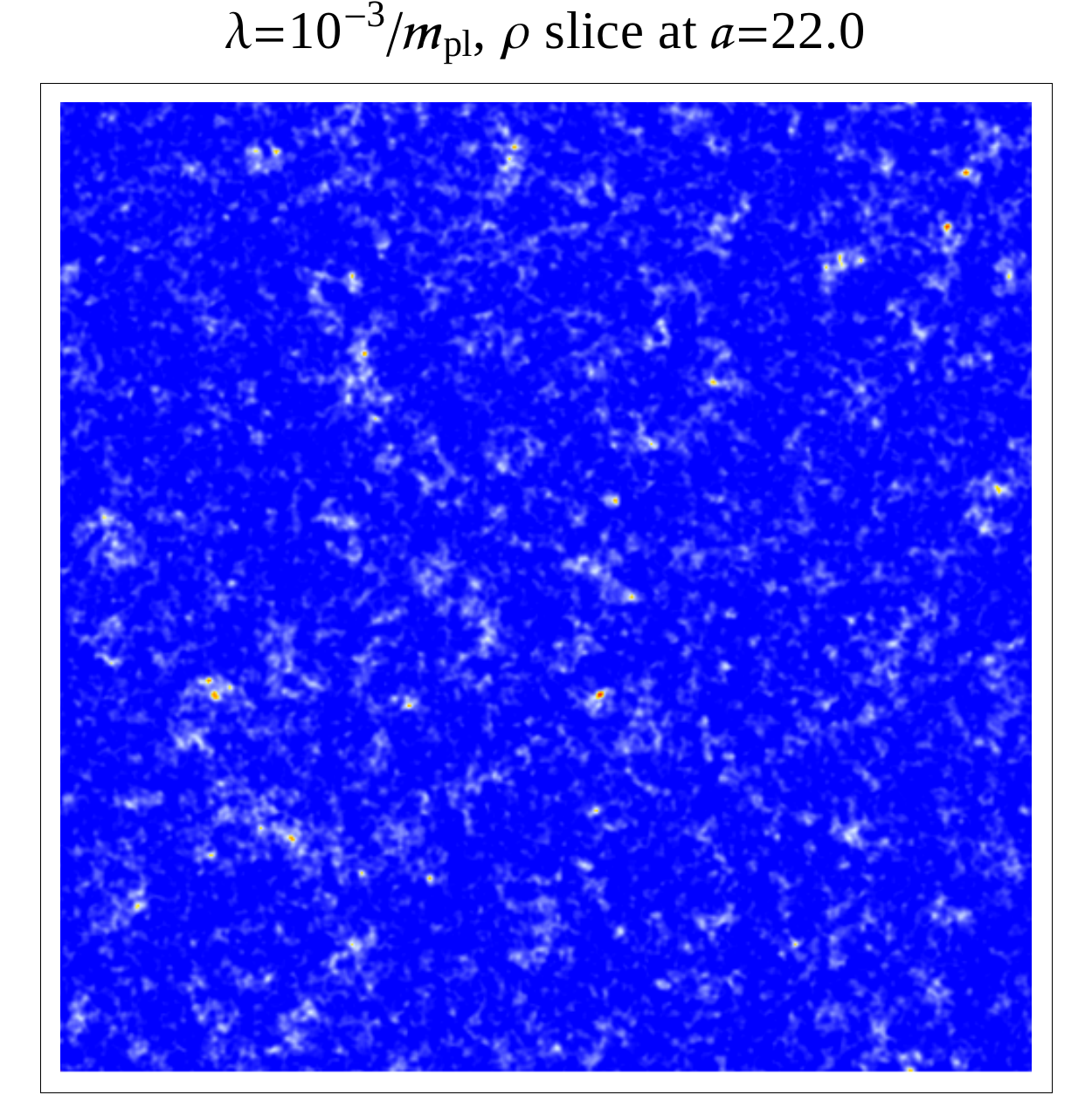}\\
\includegraphics[width=0.85\textwidth]{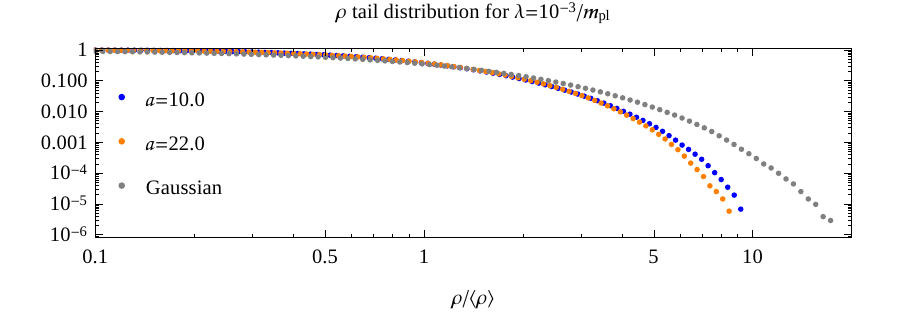}
\end{array}$
 \caption{{\it Above}: energy density slices at $a=10$ (left) and $a=22$ (right) for $v=10^{-2}\mpl$ and $\lambda=10^{-3}/\mpl$, from a simulation with $1024^2$ points. No oscillons can be seen both at $a=10$ and $a=22$. \newline
{\it Below}: tail distribution of the energy density $\rho$ as a function of $\rho/\langle \rho \rangle$ at $a=10$ (blue) and $a=22$ (orange) for $\lambda=10^{-3}/\mpl$. In grey we show the tail distribution of $\rho_g = 36V_0 g_1^2/v^2 + \lambda^2 v^4g_2^2/2$ where $g_1$ and $g_2$ are two discrete Gaussian random fields defined on a lattice with $1024^2$ points, with  zero means and standard deviations chosen such that $\langle g_1^2\rangle/\langle g_2^2 \rangle \sim \langle\delta\phi^2\rangle/\langle\delta\chi^2\rangle\sim1/5$. The tail distributions at $a=10$ and $a=22$ are both very close to the Gaussian tail distributions, indicating that no oscillons are present. }
  \label{fig:ehist3}
\end{figure}

\begin{figure}[tbp]
  \centering$
\begin{array}{cc}
\includegraphics[width=0.4\textwidth]{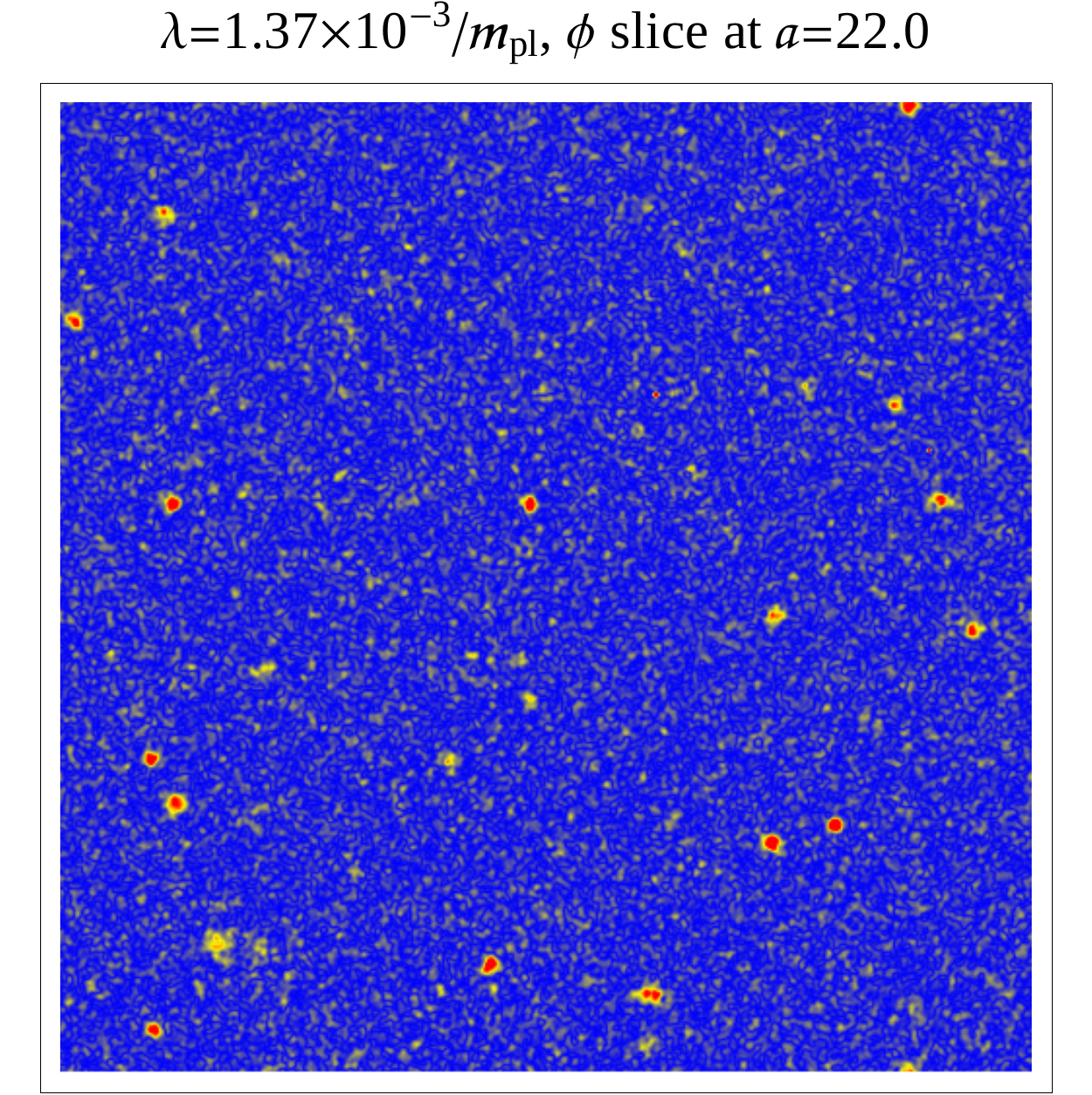}\;\;
\includegraphics[height=6.5cm]{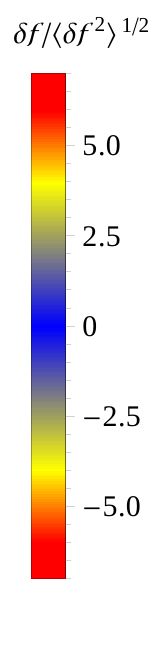}
\includegraphics[width=0.4\textwidth]{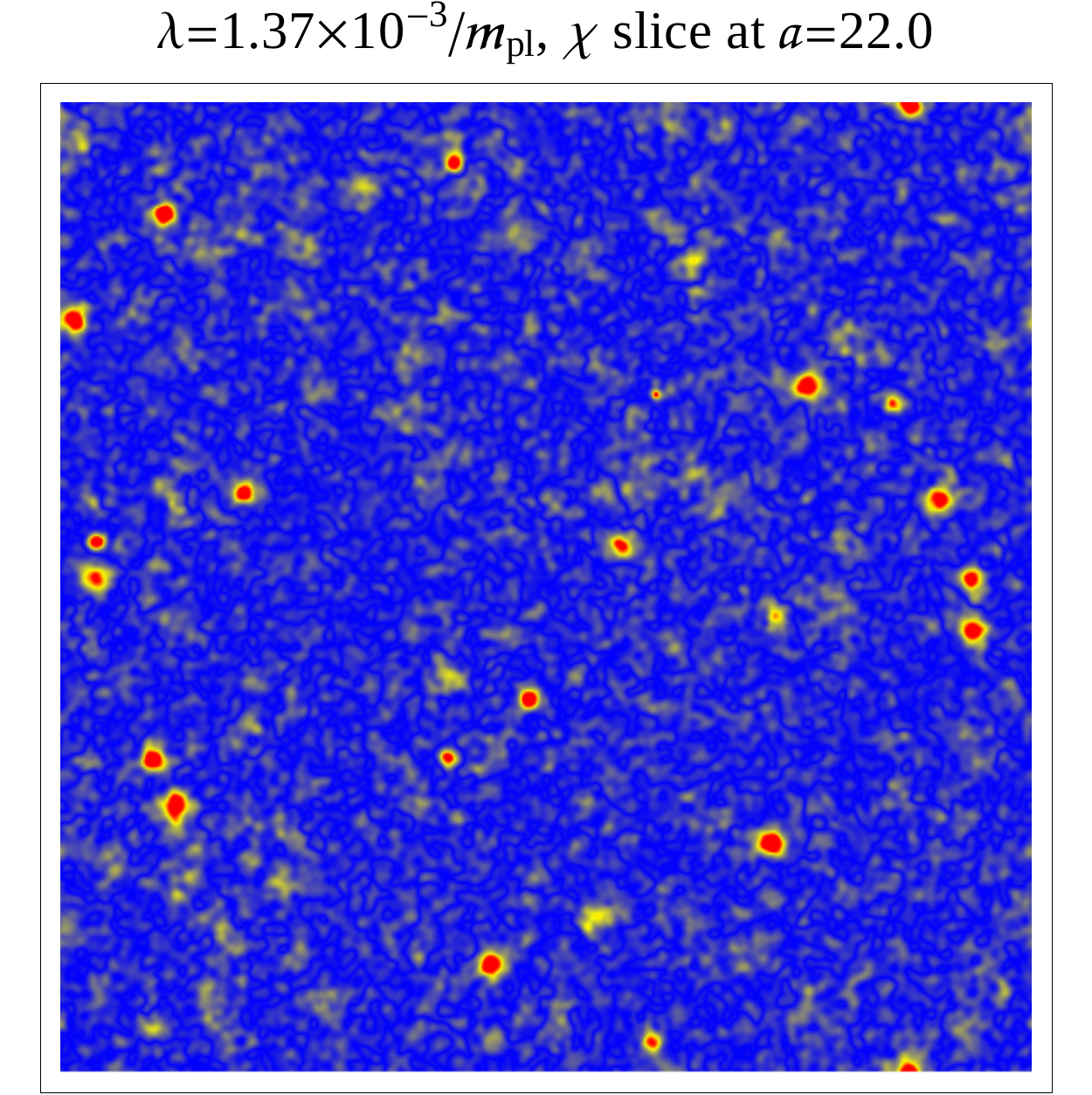}
\end{array}$
 \caption{Slices of $\phi$ (left) and $\chi$ (right) at $a=22$ for $v=10^{-2}\mpl$ and $\lambda=1.37\times10^{-3}/\mpl$, from a simulation with $1024^2$ points. The red spots in both slices show that both $\phi$ and $\chi$ constribute to the oscillons. The fluctutaions of the two fields are clearly correlated, although the $\chi$ red spots seem larger than those of $\phi$.
 Note that not all correlations may be visible since the localized oscillations in $\phi$ and $\chi$ can be out of phase. 
 }
  \label{fig:fields}
\end{figure}

In this section, we present and discuss the results of four lattice simulations in $2+1$ dimensions with $N=1024$ and other parameters, given in table \ref{tab:ini}. The simulations correspond, respectively, to $\lambda=0$ (single-field case), $\lambda=1.5\times10^{-3}/\mpl$,  $\lambda=1.37\times10^{-3}/\mpl$ and $\lambda=10^{-3}/\mpl$. The latter two cases are two-field cases where $\chi$'s fluctuations grow due to a parametric resonance caused by the inhomogeneous $\phi$ field.

We start by discussing the position space slices of the energy density $\rho$ in \ref{sssec:ps} (figs.~\ref{fig:ehist1}, \ref{fig:ehist2} and \ref{fig:ehist3}), since they provide the most intuitive illustration of the formation and evolution of oscillons. Next, in \ref{sssec:td} we discuss the energy density tail distributions $TD_\rho$ (same figures, below the $\rho$ slices). They are defined as the probability that $\rho$ is greater than a value $\varrho$, that is $TD_\rho(\varrho)=P(\rho>\varrho)$. Finally, in~\ref{sssec:evo} and fig.~\ref{fig:omega10}, we present the time evolution of the quantity $\rho_{0.1}$, which we define later and illustrates the evolution of the oscillons, emphasizing the dependence of their stability on $\lambda$.

\subsubsection{Position slices}
\label{sssec:ps}

Figs.\ \ref{fig:ehist1}, \ref{fig:ehist2} and \ref{fig:ehist3} show the energy density slices at times $a=10$ and $a=22$ , together with the corresponding energy tail distributions, for $\lambda=1.5\times10^{-3}/\mpl$, $\lambda=1.37\times10^{-3}/\mpl$ and $\lambda=10^{-3}/\mpl$, respectively. Fig.~\ref{fig:fields} shows the field values of $\phi$ and $\chi$ for slices at $a=22$ for $\lambda=1.37\times10^{-3}/\mpl$.

Let us first look at the $\rho$ slices for the coupling outside the resonance band, i.e.\ $\lambda=1.5\times10^{-3}\mpl$, in fig.~\ref{fig:ehist1}. The upper-left slice shows $\rho$ at $a=10$. We can clearly see many red spots, which correspond to overdensities with $\rho\gtrsim9\langle\rho\rangle$. These spots are localized oscillations of the inflaton field around the minimum, i.e.\ oscillons. The upper-right slice shows $\rho$ at $a=22$. By then, most of the oscillons present at $a=10$ have disappeared and just a few remain. Comparing the case $\lambda=1.5\times10^{-3}\mpl$ with the single-field case $\lambda=0$ we find that the evolution of the oscillons is qualitatively the same: a coupling $\lambda$ outside the resonance bands does not significantly affect the formation and stability of the oscillons.

For $\lambda=1.37\times10^{-3}/\mpl$, at $a=10$ the situation is similar to the non-resonant $\chi$ case. Indeed, the upper-left slice of fig. \ref{fig:ehist2} contains oscillons, analogously to the corresponding slice with $\lambda=1.5\times10^{-3}/\mpl$. On the other hand, at $a=22$ the situation is very different: one can see in the upper-right slice of fig.\ \ref{fig:ehist2} that the oscillons have not decayed. On the contrary, they seem to have expanded compared to $a=10$, i.e.\ the presence of $\chi$, enhanced by the parametric resonance, makes the oscillons more stable for $\lambda=1.37\times10^{-3}/\mpl$. The relative contributions of $\phi$ and $\chi$ to the oscillons can be seen in the field slices at $a=22$ shown in fig. \ref{fig:fields}. Both fields contribute to the oscillons, however those the localized $\chi$ fluctuations look somewhat larger. Furthermore, one can see that the fluctuations in the two fields are correlated: by looking at the two slices in fig.\ \ref{fig:fields} we can pair many red spots of the $\phi$ slice with red spots of the $\chi$ slice. It should be noted, however, that it may be that not all correlations are identified, since the oscillations of $\phi$ and $\chi$ may be out of phase at the time of shown slices and thus one of the red spots may not be visible.

Finally, the $\rho$ slices for $\lambda=10^{-3}/\mpl$ are shown in fig.\ \ref{fig:ehist3}, where a strong parametric resonance of $\chi$ happens earlier. In contrast to the other cases, no oscillons can be seen at $a=10$ and $a=22$. This is a dramatic difference to the two other cases and emphasizes the strong dependence of the fate of oscillons on the timing of the parametric resonance. Note that the fluctuations of $\chi$ start growing after the inflaton field has become inhomogeneous: the resonance happens shortly after hill crossing \cite{Antusch:2015nla} and rapidly destabilizes the oscillons. While we have seen that a delayed parametric resonance of $\chi$ (e.g.\ for $\lambda=1.37\times10^{-3}/\mpl$ discussed above) enhances the oscillons and makes them more stable, an early and strong parametric resonance of $\chi$ has the opposite effect and strongly suppresses the oscillons.

\subsubsection{Energy density tail distributions}
\label{sssec:td}

In order to quantify the contribution of the oscillons to the energy of the lattice it is useful to look at the tail distribution of the energy density of the lattice. The tail distribution $TD_\rho$ of a real random variable $\rho$ is defined as the probability that $\rho$ is greater than a value $\varrho$, that is $TD_\rho(\varrho)=P(\rho>\varrho)$.

The tail distributions of the energy density at times $a=10$ and $a=22$ can be seen in figs.~\ref{fig:ehist1}, \ref{fig:ehist2} and \ref{fig:ehist3}, for $\lambda=1.5\times10^{-3}/\mpl$, $\lambda=1.37\times10^{-3}/\mpl$ and $\lambda=10^{-3}/\mpl$, respectively. In order to compare the tail distributions to a Gaussian expectation, we generated two discrete Gaussian random fields $g_1$ and $g_2$ on a lattice with $1024^2$ points, with zero mean and standard deviations chosen such that $\langle g_1^2\rangle/\langle g_2^2 \rangle \sim \langle\delta\phi^2\rangle/\langle\delta\chi^2\rangle$. We then define $\rho_g = 36V_0 g_1^2/v^2 + \lambda^2 v^4g_2^2/2$ and compare its tail distribution to the ones extracted from the lattice simulations with corresponding $\lambda$. Of course, we do not expect the fields to be exactly Gaussian. It is nevertheless useful to compare the lattice distributions to the ones for  Gaussian random fields because the latter provide examples of field configurations with same variances but clearly without oscillons. The comparison allows to distinguish the overdensities from the expected statistical fluctuations of the energy density.

In the $\lambda=1.5\times10^{-3}/\mpl$ case, shown in fig.~\ref{fig:ehist1}, at $a=10$ the energy density tail distribution reaches values of $\rho>300\langle\rho\rangle$.  This is what one would expect from a lattice containing many oscillons, with a large fraction of the energy density contained in regions with $\rho\gg\langle\rho\rangle$. By $a=22$, the tail distribution has shrunk, closer to the expectation for a Gaussian energy density $\rho_g$. For $\lambda=0$, we obtained the same results. This means that the oscillons start decaying within our simulations when $\lambda$ lies outside the resonance band (and for $\lambda=0$), as we can also see from the energy density slices. 

For $\lambda=1.37\times10^{-3}/\mpl$, see fig.~\ref{fig:ehist2}, both at $a=10$ and $a=22$ the tail distributions reach values of order $1000\langle\rho\rangle$, much larger than the Gaussian expectation. This confirms what we see in the energy density slices: the oscillons do not decay within the simulation when $\lambda=1.37\times10^{-3}/\mpl$. In this case, the parametric resonance of $\chi$, sourced by the inhomogeneous $\phi$, stabilizes the oscillons.

Lastly, fig.~\ref{fig:ehist3} shows the tail distributions for $\lambda=10^{-3}/\mpl$. Here, the tail distributions at $a=10$ and $a=22$ are both very close to the Gaussian expectation. The oscillons have already decayed at $a=10$: in this case, the parametric resonance of $\chi$ suppresses the oscillons.

\subsubsection{Evolution of the energy density}
\label{sssec:evo}

\begin{figure}[tbp]
  \centering$
\begin{array}{cc}
\includegraphics[width=0.9\textwidth]{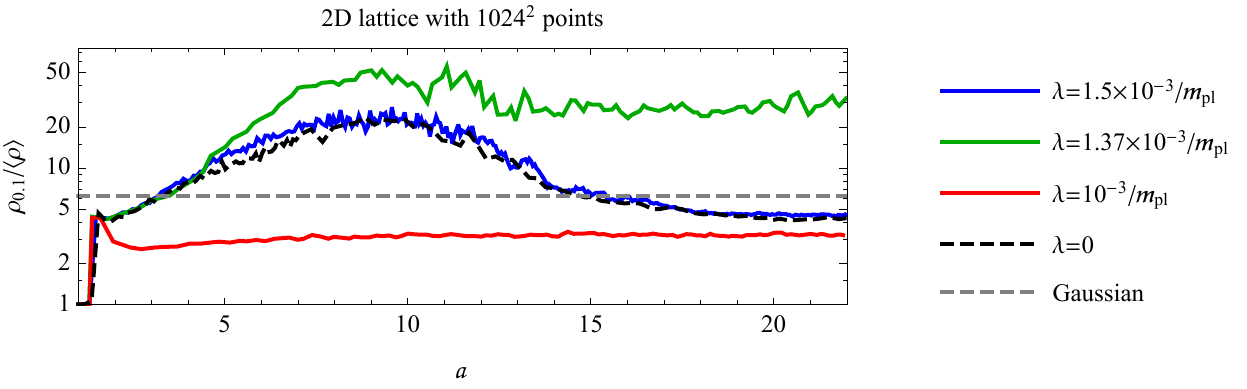}
\end{array}$
 \caption{The value $\rho_{0.1}$ for which $10\%$ of the energy resides regions with $\rho > \rho_{0.1}$ for $v=10^{-2}\mpl$ and $\lambda=1.5\times10^{-3}/\mpl$ (blue), $\lambda=1.37\times10^{-3}/\mpl$ (green), $\lambda=10^{-3}/\mpl$ (red) and $\lambda=0$ (dashed black). The gray dashed line is what one would expect for $\rho_g\propto g^2$, where $g$ is a discrete Gaussian random field with zero mean, defined on a lattice with $1024^2$ points. The formation of oscillons drives $\rho_{0.1}/\langle\rho\rangle$ to larger values. For $\lambda=1.5\times10^{-3}/\mpl$ and $\lambda=0$, the lines track each other, showing that a coupling outside the resonance band has no effect on the evolution of oscillons. For $\lambda=1.37\times10^{-3}/\mpl$ one can see that $\rho_{0.1}/\langle\rho\rangle$ increases at the same pace as the single-field and non-resonant cases until $a\sim4$. Afterwards, the growth of the fluctuations of $\chi$ drives $\rho_{0.1}/\langle\rho\rangle$ to larger values for $\lambda=1.37\times10^{-3}/\mpl$, reaching a maximum of $\sim50$ at $a\sim10$ before settling at $\sim40$. On the other hand, for $\lambda=1.5\times10^{-3}/\mpl$ and $\lambda=0$, $\rho_{0.1}/\langle\rho\rangle$ decreases down to $\sim4$ after reaching its maximum of $\sim20$ at $a\sim10$, indicating that oscillons decay in the single-field and non-resonant cases in $2+1$ dimensions. Finally, for $\lambda=10^{-3}/\mpl$, $\rho_{0.1}/\langle\rho\rangle$ initially grows to values close to $5$, corresponding to the initial stages of oscillon formation. However, it subsequently drops to lower values, indicating that the oscillons are strongly suppressed in this case. }
  \label{fig:omega10}
\end{figure}

Fig.~\ref{fig:omega10} summarizes the results discussed in the previous subsections. It shows the value $\rho_{0.1}$ for which $10\%$ of the energy is contained in regions with energy density $\rho > \rho_{0.1}$ for the four simulations. That is, $\rho_{0.1}$ is defined by:

\be
\sum_{\substack{\bar{x}\;{\rm with} \\ \rho(\bar{x})>\rho_{0.1}}} \rho(\bar{x}) = 0.1 \sum_{{\rm all}\;\bar{x}} \rho(\bar{x})\,,
\en
With oscillons being localized overdensities of energy, larger values of $\rho_{0.1}$ indicate more (or more spatially extended, or more energetic) oscillons. 

The curves for $\lambda=1.5\times10^{-3}/\mpl$ (blue in the figure) and $\lambda=0$ (dashed black) track each other, emphasizing that a coupling outside the resonance band has no effect on the formation and stability of the oscillons. In both cases, one can see that $\rho_{0.1}$ reaches $\sim20\langle \rho \rangle$ at $a\sim10$ before decreasing to $\sim4\langle \rho \rangle$, indicating that most oscillons decay by the end of the simulation. Therefore, a coupling outside the resonance band has no effect on the oscillons.

For $\lambda = 1.37\times10^{-3}/\mpl$ (green), where $\chi$'s fluctuations are amplified at $a\sim4$, $\rho_{0.1}$ reaches $\sim50\langle \rho \rangle$ at $a\sim10$ before settling at $\sim40\langle \rho \rangle$. In this case, the parametric resonance of $\chi$ enhances the oscillons and stabilizes them.

For $\lambda = 10^{-3}/\mpl$ (red), where the parametric resonance of $\chi$ occurs at $a\sim2$, $\rho_{0.1}$ rapidly settles at $\sim3\langle \rho \rangle$, close to the Gaussian expectation (dashed gray). In contrast to the $\lambda=1.37\times10^{-3}/\mpl$ case, the formation of oscillons is strongly suppressed by $\chi$.

The factor that determines whether the growth of the fluctuations of $\chi$ enhances or suppresses the formation of oscillons is the time at which the growth starts. Indeed, as can be seen in fig.~\ref{fig:vars}, for $\lambda=1.37\times10^{-3}/\mpl$, $\langle\delta\chi^2\rangle$ starts growing around $a\sim4$, when the $\phi$ oscillons have already oscillated around $\phi=v$ many times. Fig.~\ref{fig:omega10} shows this clearly: until $a\sim4$, the evolution of $\rho_{0.1}$ for $\lambda=1.37\times10^{-3}/\mpl$ is practically the same as for the single-field case $\lambda = 0$ and for $\lambda=1.5\times10^{-3}/\mpl$, where $\chi$ is not amplified. The contribution of the $\phi$ oscillons to the energy density until $a\sim4$ is the same in the three cases. After this phase, when $\langle\delta\chi^2\rangle\sim\langle\delta\phi^2\rangle$, the two curves diverge and, for $\lambda=1.37\times10^{-3}/\mpl$, the oscillons' contribution to the energy density is enhanced. As can be seen in fig.~\ref{fig:fields}, in this case also $\chi$ forms oscillons. 

On the other hand, for $\lambda=10^{-3}/\mpl$, the growth of $\langle\delta\chi^2\rangle$ starts earlier, just after the hill crossing and during the formation phase of the oscillons: already at $a=2$, $\langle\delta\chi^2\rangle\gtrsim\langle\delta\phi^2\rangle$. Correspondingly, the value of $\rho_{0.1}$ decreases and the formation of oscillons is suppressed.

\section{Conclusions}
\label{sec:conc}

In this paper we have investigated how oscillons, which are spatially localized and relatively stable fluctuations of the inflaton field $\phi$ that can form after inflation, are affected by the couplings to other scalar fields. We considered hilltop inflation models of the type studied in \cite{Antusch:2014qqa,Antusch:2015nla,inprep}, where we have recently shown that hill crossing oscillons generically form \cite{Antusch:2015nla}. Furthermore, in \cite{inprep} we have shown that when another scalar field $\chi$ is coupled to the inflaton, depending on the value of the coupling parameter, the latter can get resonantly enhanced by the inhomogeneous inflaton field.

In order to study the effect of $\chi$ on the oscillons, we solved the discretized equations of motion on a lattice with $2+1$ dimensions for different values of the coupling constant, including the single-field case (which corresponds to zero coupling). The other choices of the coupling constant $\lambda$ were guided by the observations of \cite{inprep}: we chose one value for which the parametric resonance of $\chi$ happens around the time at which oscillons form, $\lambda=10^{-3}/\mpl$, a second value for which it happens later and is somewhat less efficient, $\lambda=1.37\times10^{-3}/\mpl$, and a third value $\lambda=1.5\times10^{-3}/\mpl$ for which no resonant enhancement occurs.   

In contrast to earlier studies \cite{Gleiser:2014ipa} which used a similar potential and observed no effects of other scalar fields on the oscillons, we found that the impact of the coupling to $\chi$ can be very strong and that three cases can be realized: 
for a fast and strong parametric resonance of $\chi$ (e.g.\ for $\lambda=10^{-3}/\mpl$), oscillons are strongly suppressed. 
For a delayed and somewhat weaker parametric resonance (e.g.\ for $\lambda=1.37\times10^{-3}/\mpl$), the inflaton oscillons get imprinted on the other scalar field and their stability is even enhanced compared to the single-field oscillons.
Only when no parametric resonance occurs (e.g.\ for $\lambda=1.5\times10^{-3}/\mpl$), and the $\chi$ field stays subdominant, the additional scalar field has no effects on the oscillons.  We expect the resonant amplification of $\chi$ caused by the inhomogeneous inflaton to also have a strong impact on oscillons in $3+1$ dimensions.

Of course, the other scalar field can be excited for other reasons than parametric resonance, i.e.\ it might have an initially large homogeneous mode or could be produced by fast perturbative decays. In such cases, we would also expect potentially strong effects on oscillons. However, the study of such scenarios is beyond the scope of this paper.

Another interesting question that we did not address, is the effect of quantum fluctuations on the evolution of oscillons after hilltop inflation. It has been shown in \cite{Saffin:2014yka} that couplings to fermions can decrease the classical lifetime of oscillons. This question needs further investigation, although, since the effect of the other scalar field starts relatively early in the oscillons' evolution, we expect that the enhancement and suppression of oscillons we observed will remain even when quantum corrections are taken into account.

In summary, we found that the couplings of the inflaton to other scalar fields can have a strong impact on the formation and stability of oscillons, especially when the latter fields are resonantly enhanced. Such impact affects the expansion history of the universe, possibly leading to observable effects from oscillons after inflation.

\section*{Acknowledgements}

This work has been supported by the Swiss National Science Foundation.

\end{document}